# Green Hydrogen Cost-Potentials for Global Trade

D. Franzmann,[1,2,*] H. Heinrichs,[1] F. Lippkau,[4] T. Addanki,[3] C. Winkler,[1,2] P. Buchenberg,[3] T. Hamacher,[3] M. Blesl,[4] J. Linßen,[1] and D. Stolten[1,2]

1 Forschungszentrum Jülich GmbH, Institute of Energy and Climate Research – Techno-economic Systems Analysis (IEK-3), 52425 Jülich, Germany

2 RWTH Aachen University, Chair for Fuel Cells, Faculty of Mechanical Engineering, 52062 Aachen, German

3 Chair of Renewable and Sustainable Energy Systems - TU Munich; ens@ei.tum.de

4 Institute of Energy Economics and Rational Energy Use (IER), University of Stuttgart, 70565 Stuttgart, Germany

* corresponding author: d.franzmann@fz-juelich.de

## Abstract

Green hydrogen is expected to be traded globally in future greenhouse gas neutral energy systems. However, there is still a lack of temporally- and spatially-explicit cost-potentials for green hydrogen considering the full process chain, which are necessary for creating effective global strategies. Therefore, this study provides such detailed cost-potential-curves for 28 selected countries worldwide until 2050, using an optimizing energy systems approach based on open-field photovoltaics (PV) and onshore wind. The results reveal huge hydrogen potentials (>1,500 $PWh_{LHV}$/a) and 79 $PWh_{LHV}$/a at costs below 2.30 EUR/kg in 2050, dominated by solar-rich countries in Africa and the Middle East. Decentralized PV-based hydrogen production, even in wind-rich countries, is always preferred. Supplying sustainable water for hydrogen production is needed while having minor impact on hydrogen cost. Additional costs for imports from democratic regions are only total 7% higher. Hence, such regions could boost the geostrategic security of supply for greenhouse gas neutral energy systems.

**Keywords (max 6):** Green hydrogen, cost-potentials, LH2, greenhouse gas-neutral, energy system, energy security

## 1 Introduction

Green hydrogen has been advocated by various studies as a key element in the transformation of the energy system towards greenhouse gas-neutrality [1]. The main reasons for this are that it can serve as seasonal bulk storage to counteract the volatility of wind and solar energy technologies [2] and can be used as a material to support the decarbonization of challenging sectors like the chemical industry [3]. Although hydrogen can be produced in various ways, green hydrogen produced via water electrolysis powered by renewable energy technologies is preferred in most scenarios thanks to its minimal carbon footprint [4]. Hence, the regional conditions for renewable energy sources (RES) determine a large portion of the achievable hydrogen production costs [5], [6]. As wind speeds, solar irradiation, and suitable locations for both renewable sources vary largely across the world, the idea arose to make use of this difference by trading green hydrogen internationally [7]. In this context, different transport options for green hydrogen are currently being discussed [8] and various bilateral agreements have been signed or are under consideration [9]. Against this background, knowledge about spatially-resolved costs and potentials for green hydrogen production as a potential globally-traded energy carrier is mandatory.

However, recent studies have primarily focused on the local or national production of hydrogen [10],[11], [12], such as for Niger in Bhandari [13] or Turkey in Karayel et al. [14]. On the other hand, most global studies focus on electricity potentials from renewable energy sources, rather than considering the potential for hydrogen export or import [15, 16]. Only a few studies have investigated the potentials for green hydrogen on a global scale as well for carriers based on hydrogen. One common approach is to divide the world into evenly spaced gridded cells based on latitude and longitude coordinates and to calculate green hydrogen cost based on cost optimizations for each grid cell [6][17][18].The IRENA Green Hydrogen Report [6] calculates the costs and potentials for green hydrogen globally, but does not take into account the total system necessary for exporting it. Instead, its approach only considers time series for the sizing of photovoltaics (PV), onshore wind, and electrolysis for 1x1km² stand-alone systems. The study of Sens et. al. [18] uses a similar approach based on grid cells to calculate country specific hydrogen potentials within Europe and the North African / MENA region as well as import costs for gaseous hydrogen transport to Germany via pipeline. They find that North Africa can provide gaseous green hydrogen below 2 $EUR/kg_{H2}$ in 2050 based on hybrid generation from onshore wind and PV. Also, they notice a high curtailment rate of the renewable energy production within the energy system in general, which drops substantially as soon as large scale hydrogen storage like salt caverns become available resulting in parallel to decreasing hydrogen cost . Fasihi et. al [17] uses the grid cell approach for a global evaluation of baseload hydrogen production. For each 0.45°x0.45° cell, a cost-optimal energy system including batteries and underground salt cavern hydrogen storages are utilized, leading to gaseous hydrogen cost of 1.18 EUR/kg in 2050 in Africa and South America. Though, the energy system does not include costs for transportation and effects from spatial compensation via grids within the countries. A similar approach from Fasihi [19] is applied for calculating global ammonia production costs. All gridded approaches neglect synergies with spatial compensations via grid and often temporal compensations via storage units. In addition, the approach does not consider costs for the transportation of hydrogen to a location of export.

In contrast to the gridded approach, a study from Janssen et. al. [20] shows cost prognoses for different European countries. The expected costs for gaseous hydrogen in 2050 are as low as 1.7 EUR/kg in 2050 in Ireland. Yet, the approach does not consider energy potentials for RES and hydrogen, as only average capacity factors for onshore wind and PV are applied at each country. Brändle [21] proposes a green hydrogen import cost tool, and considers a similar country wise approach as Janssen et. al.[20] . He utilizes a cost optimization of generation and electrolysis units, which is based on average capacity factors and synthetic time series optimizing the sizing country wide. Therefore, this approach does not take the entire hydrogen export energy system into account and , hence, neglects possible impacts of storage and grid-balancing needs. The approach was also used by Moritz et. al [22] to calculate export costs for other hydrogen based energy carriers. Buchenberg et. al. [23] considers a more complex energy system designs for synfuel exports, but does not model spatially resolved transportation. Also, studies based on integrated assessment models exist like a study from van der Zwaan et. al [24], which shows an positive economic impact of energy imports on the European and North African energy system. Yet, these studies lack a detailed spatial and temporal resolution.

Hence, there remains a gap in knowledge regarding the impact of the full green hydrogen process chain design on hydrogen production costs for export needs. Therefore, this paper attempts to close this gap by investigating highly temporally- and spatially-resolved green hydrogen production costs in countries around the world, including the full infrastructure necessary for exporting liquid hydrogen. By this, the impact of the full export infrastructure on cost distributions across a selected set of suitable countries will be revealed. Liquid hydrogen is chosen as the exported energy carrier for hydrogen, as it is expected that this option will result in the lowest shipping costs in the long term [5]. Also, it has the least environmental impacts [25–27]. The latter is especially advantageous, as liquid hydrogen is not toxic, in contrast to, for example, ammonia [28]. As renewable energy sources, wind turbines and open-field photovoltaics (PV) are selected, as both are expected to play a crucial role in future energy systems due to their comparably low costs and abundant global potential [1, 15, 29]. In addition, they avoid some of the severe environmental impacts that can occur with renewable energy sources such as hydropower and bioenergy [30–32]. The obtained information regarding potentials and full infrastructure costs for green hydrogen exports around the world can serve as a basis for developing import strategies for various countries and, hence, is provided as open data in the Appendix of this paper.

For this purpose, the utilized methodology is shown in full detail in Section 2. In Section 3, the obtained results are presented, and a special focus is given to the distribution of global green hydrogen potentials and energy system characteristics that support low hydrogen production costs. In Section 4, the paper is complemented by a discussion of the obtained results and conclusions are drawn therefrom.

## 2 Methodology

For this study, an approach is utilized that calculates the liquid hydrogen export cost based on volatile renewable energy sources for a given set of countries, including all process steps along the production chain. The goal is to calculate the export cost-potential curves for countries with the most beneficial renewable energy potentials across the world to determine differences in export cost to serve as a basis for developing strategies for reliable green hydrogen imports.

An overview of the applied approach is shown in Figure 1. It consists of five steps along the process chain, starting with volatile renewable energy sources and ending within an export harbor. Each step is described in more detail in the upcoming subsections. In the first step, countries with high solar and wind resources are selected. For these, a land eligibility analysis is conducted. Based on the available areas for renewable energy technologies, the electricity time series of these renewable energy technologies are simulated hourly from Merra-2 weather data for open-field PV and onshore wind. Afterwards, the results are clustered to serve as an input for the energy system optimizations to design the respective hydrogen export infrastructures.

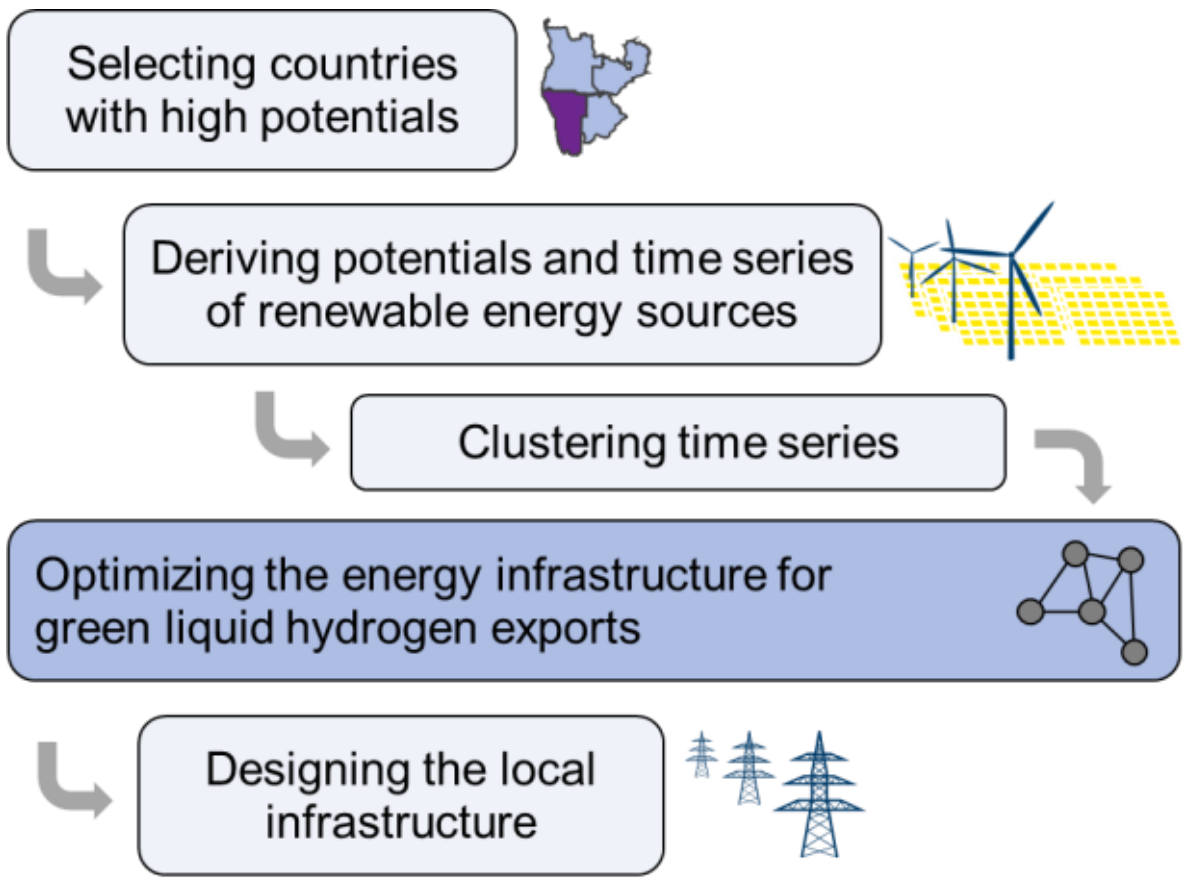

**Figure 1. Visualization of the main steps for calculating hydrogen export costs.**

To determine the cost-potential curves for liquid hydrogen exports, for each considered country 36 discrete energy systems are optimized with nine varying expansion degrees of renewable energy technologies for four different years, ranging from 2020 to 2050. The chosen approach is based on the energy system framework FINE [33, 34] and uses a holistic optimization to minimize the total system costs, which is equal to the liquid hydrogen costs in the harbor. Subsequent to the energy system optimization, the utilized renewable energy potentials are allocated to specific locations and clustered into parks. The parks are connected by designing a local transmission grid for the hydrogen export to account for sub-regional infrastructure costs as well. Finally, the obtained single cost-potentials for liquid hydrogen exports are merged into full cost-potential curves for each exporting harbor.

## 2.1 Deriving the potentials of volatile renewable energy sources

The selection of the considered countries for renewable energy potentials was based on the Global Wind Atlas [35] for wind energy and on the Global Solar Atlas [36] for solar energy. This analysis focuses on wind turbines and open-field PV, as both are perceived to be cornerstones of a global energy transformation [1, 37] and bear the comparable lowest environmental and social impacts amongst renewable energy technologies [38]. To achieve a good and well-balanced global coverage, the country selection is based on the world regions utilized in the global energy system model TIAM [39], which aims to represent the full global energy system and transformation pathways most properly. Hence, for each region in TIAM, at least one promising country in terms of volatile renewable energy potentials is selected. This results in 28 selected countries, as shown in Figure 2.

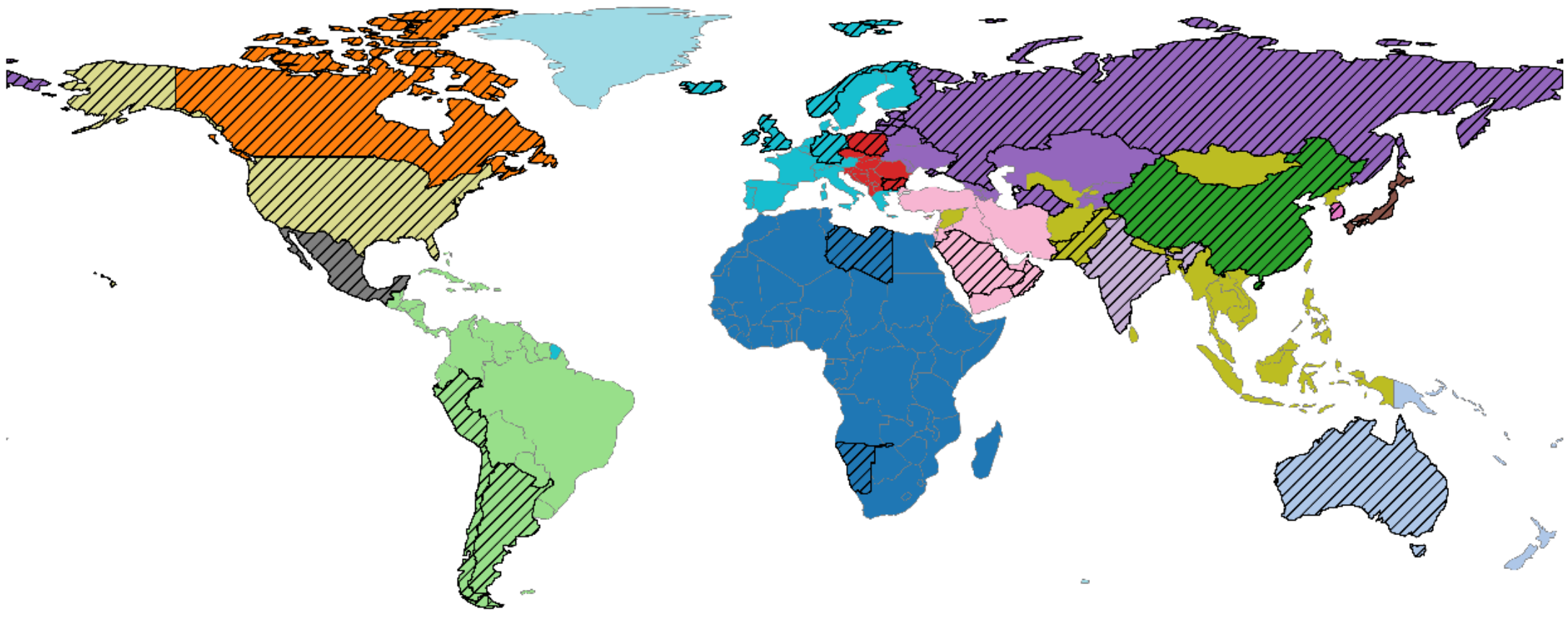

**Figure 2. TIAM regions (colored) and selected countries (striped).**

For the technical electricity potentials of open-field PV and onshore wind turbines, results calculated using the open-source tool pyGRETA [40] from Buchenberg et al. [23] are utilized. Those potentials are derived by converting historical weather data for 2019 from MERRA-2 [2] and the Global Wind Atlas [35] at any location into hourly capacity factors of electricity generation [40]. This results in time series with a spatial resolution at the equator of 250 m x 250 m for wind energy and 50 km x 50 km for solar energy allocated to 250m x 250m cells for those areas, which are assumed to be suitable or available for the installation of renewable energy technologies. The applied identification of suitable areas is based on a land eligibility approach by Ryberg et al. [41]. For this study, 38 different exclusion types for land areas with specific buffer distances for each generation technology are considered (see Table 5). The exclusions consist of physical limitations like slopes, sociopolitical constraints such as distances to settlements, land use conflict with croplands and natural conservation. Finally, this results in the total technical electricity potential of open-field PV and onshore wind turbines for the considered countries as listed in Table 1 and described in more detail in Buchenberg et al. [23]. These renewable energy potential data serve as the energy source within the hydrogen export model. As the model bears a spatial aggregation at the GID-1 level (see section 2.2), the energy potentials and their time series are clustered. All available placements for renewable energy technologies within a GID-1 region are split into 11 clusters each for open-field PV and onshore wind based on their full load hours. These quantiles split the full potentials per region first into ten evenly spaced renewable energy potentials in terms of their full load hours, with an additional cluster being added representing the best 5% of the potentials to account for more details amongst the best potentials. Finally, this results in 13,376 overall clusters across the 28 considered countries, describing the total technical potential for each country.

**Table 1. Aggregated country potentials for open-field PV and onshore wind [23].**

| TIAM region | Open-field PV | | Onshore Wind | |
|---|---|---|---|---|
| | Capacity [$TW_{el}$] | Energy [$PWh_{el}$] | Capacity [$TW_{el}$] | Energy [$PWh_{el}$] |
| Libya | 139.49 | 292.93 | 10.45 | 15.41 |
| Namibia | 16.32 | 35.75 | 3.22 | 4.33 |
| Australia | 350.32 | 734.41 | 36.61 | 67.93 |
| Canada | 109.45 | 96.25 | 39.66 | 46.16 |
| China | 137.74 | 240.15 | 40.99 | 45.89 |
| Argentina | 102.49 | 177.90 | 14.39 | 26.84 |
| Chile | 22.49 | 48.75 | 2.83 | 2.48 |
| Peru | 12.19 | 27.28 | 5.10 | 0.74 |
| Germany | 0.34 | 0.39 | 0.28 | 0.56 |
| Bulgaria | 0.14 | 0.22 | 0.28 | 0.19 |
| Poland | 0.26 | 0.32 | 0.41 | 0.86 |
| Estonia | 0.12 | 0.13 | 0.16 | 0.37 |
| Lithuania | 0.12 | 0.13 | 0.21 | 0.29 |
| Latvia | 0.18 | 0.20 | 0.26 | 0.44 |
| Russia | 88.95 | 73.84 | 82.49 | 80.54 |
| Turkmenistan | 29.44 | 50.25 | 2.79 | 3.75 |
| India | 17.41 | 31.61 | 11.12 | 8.69 |
| Japan | 0.27 | 0.40 | 0.80 | 0.83 |
| South Korea | 0.05 | 0.08 | 0.18 | 0.19 |
| Oman | 25.00 | 52.11 | 1.86 | 2.00 |
| Saudi Arabia | 141.95 | 294.59 | 10.67 | 13.8 |
| Mexico | 47.24 | 93.20 | 8.84 | 6.1 |
| Pakistan | 21.37 | 42.25 | 3.59 | 3.03 |
| USA | 111.87 | 188.40 | 35.49 | 49.29 |
| United Kingdom | 0.60 | 0.61 | 0.33 | 0.85 |
| Ireland | 0.14 | 0.15 | 0.15 | 0.42 |
| Iceland | 1.62 | 1.29 | 0.35 | 0.9 |
| Norway | 2.39 | 1.92 | 0.97 | 0.94 |

### 2.2 Designing the infrastructure for green liquid hydrogen exports

The aim of the liquid hydrogen export models is to find the cost-optimal solution to convert renewable electricity from onshore wind and open-field PV to liquid hydrogen at an export harbor. Several types of carriers for transporting hydrogen like gaseous hydrogen (GH2), ammonia (NH3), or liquid organic hydrogen carriers (LOHCs) can be considered [42–44]. Based on a study from Heuser et al. [5], liquid hydrogen ($LH_2$) is chosen as the hydrogen carrier for exporting hydrogen in future global hydrogen markets for this study, due to its low shipping and reconversion cost in the long term. Liquid hydrogen needs most part of its energy for conversion at the location of export, where the process can utilize low cost renewable energy of the exporting country, whereas ammonia and LOHC both have a high demand for high temperature heat occurring within the importing country, which imports hydrogen or hydrogen carrier mostly due to limits in local renewable energy expansion at sufficiently low cost. Hence, this could result in a barrier for ammonia and LOHC depending on the use in the importing country [6]. Additionally, liquid hydrogen has higher conversion efficiencies and lower investment costs compared to LOHC and ammonia [45].

The entire structure of the underlying energy system model is shown in Figure 3. It consists of the renewable energy sources on the left. The generated electricity can be converted via PEM-electrolysis and distributed via hydrogen pipelines (hydrogen path) or can also be stored in batteries and transported within an electrical grid to a harbor to supply liquefaction (electricity path). In the harbor, the hydrogen is liquefied and stored to be readied for the export of green hydrogen.

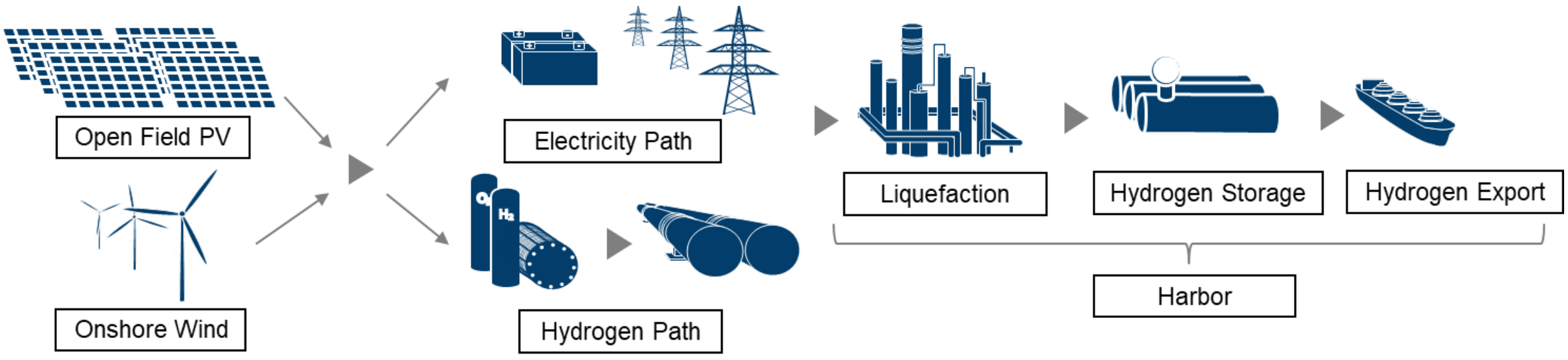

**Figure 3. Overview of the elements of the energy system model.**

As the spatial resolution of the model highly impacts the generation and infrastructure in terms of cost and design [46], each considered country is modeled on a GID-1 level, which roughly equates to federal states and varies between 85 regions for Russia and four for Great Britain.

A higher spatial resolution would result in excessively challenging computational efforts. The respective export port for each country is considered as an additional region within the model. The assumed hydrogen demand for export is allocated within this separate region. In all countries apart from Turkmenistan, the industrial port with the highest freight handling capacity based on data from marineinsight [47] and worldshipping [48] is assumed as the export location. For Turkmenistan, which has no access to an ocean for global shipping, the Pre-Caspian gas pipeline highlighted in Balkanabat [49] is assumed as the point of export. For all points of export, no capacity restrictions are applied.

The onshore wind turbines and open-field PV are modeled as electricity sources within the model. Their temporally-resolved maximum feed-in is determined by the hourly time series from the renewable energy potential analysis (see Section 2.1). As the time series, especially for wind energy, vary strongly even within GID-1 sub-regions, eleven clusters for both technologies based on their full load hours are considered within each of these. Based on these clusters, the optimization can choose which to utilize first from an energy systems perspective.

The electricity of the renewable energy technologies is partially converted into green hydrogen by means of PEM electrolysis within the region of the renewable electricity generation (decentralized) or at the export location (centralized). As the aim of this study is to derive the maximum technical liquid hydrogen potential associated with its costs based on renewable potentials, restrictions for ramp up and material use etc. are not considered and the model is allowed to expand electrolysis infinitely if required. The water usage for the hydrogen production is assumed to be from groundwater and costs are therefore already included in the techno-economic assumptions (see Table 2). In the future, groundwater will most probably not be available as an unlimited resource [50]. Yet, the additional cost of seawater desalination on hydrogen production is marginal, as Yates et al. [51] and Heinrichs et. al. [52] showed. For the grid, the model can choose between an AC electric grid and gaseous hydrogen pipelines (see Table 2). Based on the fact that in most cases, decentralized hydrogen production and transport via hydrogen pipeline is expected to be the comparably cheaper option in accordance with Reuß et al. [53], the electric grid primarily serves the purpose of supplying the electricity demand of the liquefaction of about 0.205 $kWh_{el}/kWh_{H2,LHV}$ [53] within the harbors. Both grids are modeled using a greenfield approach due to the vast required infrastructure expansions, allowing grid connections between neighboring sub-regions from centroid to centroid with a detour factor of 1.3 [54] and connecting remote parts to the nearest locations on the mainland.

The hydrogen is liquefied and stored in the export location. As was shown in Reuß et al. [53], the cost of liquefaction greatly depends on the system size. Currently, there are only plants available at a capacity below 100 tons per day [44, 55, 56]. For the large-scale applications in this study, higher capacities are needed. Therefore, the investment costs are modeled as a function of the liquefaction plant size (see Table 2). As there are currently no large-scale liquefaction plants available, the limitation of scaling effect is set as the largest LNG liquefaction train at about 20,000 tons per day [57]. As hydrogen export via ship is assumed to be constant throughout the entire year, the storage is needed to account for fluctuations in hydrogen generation.

The hydrogen energy system model is formulated and solved using the open energy system model framework FINE [33, 34] as a holistic linear optimization, taking into account all of the design and operation of the export process chain in one approach. The optimization itself is carried out with the gurobi solver [58], and is solved for one year. As the model itself only considers spatial differences for transport on the GID-1 level, all transport costs below that spatial resolution are not accounted for within the optimization. Therefore, this is considered through a post-processing step following the optimization. In this step, the local transmission grid to connect the wind turbines and PV modules first to the parks and second to the electrolysis units is calculated. This is done by deriving the actual used placements by choosing the best placements from the potentials in terms of full load hours until the capacity from the optimization is reached. All placements within a 5 km radius are clustered to wind or PV parks. Subsequently, each park is connected to the electrical grid on the GID-1 level obtained within the optimization using a minimum spanning tree [59]. Figure 4 depicts this step for a random example in which all parks are connected to the electrolysis units of the respective GID-1 region. Finally, the cost for the local transmission grid is calculated assuming the parameter from Table 2. The resulting hydrogen export costs are calculated as: $c_{H_2} = \frac{TAC_{opt}+C_{local\,grid}}{m_{H_2,export}}$ where $TAC_{opt}$ is the total annual energy system costs from the optimization, $C_{local\,grid}$ the cost for the local transmission grids, and $m_{H_2,export}$ the exported amount of hydrogen for the simulated year.

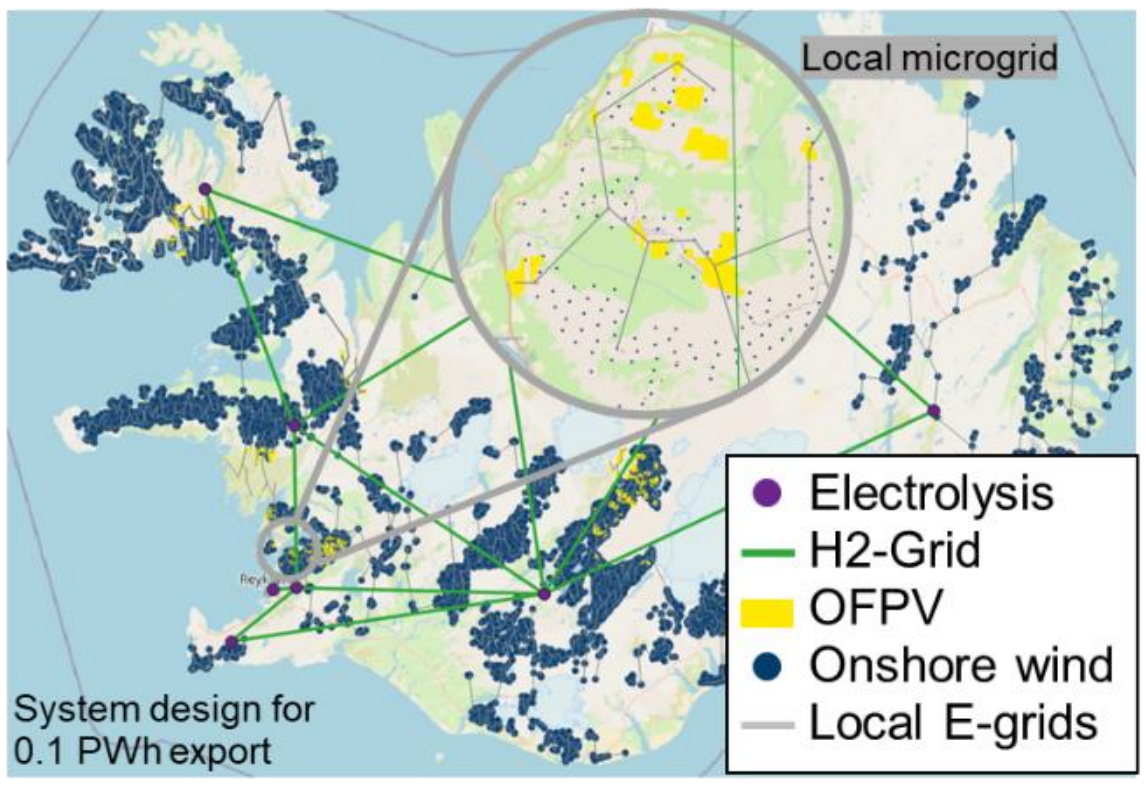

**Figure 4. Geospatial design of the energy system, including local transmission grids and the export harbor.**

**Table 2. Assumed techno-economic parameters for the considered green hydrogen export chain process steps. EUR refers to $EUR_{2022}$. The cost assumptions are calculated as total annual costs with an interest rate of 8%.**

| | CAPEX [EUR/kW] | | | | OPEX[% capex/a] | Efficiency [%] | Lifetime [years] | Source |
|---|---|---|---|---|---|---|---|---|
| | 2020 | 2030 | 2040 | 2050 | | | | |
| **Onshore wind** | 1,257 | 1,137 | 987 | 923 | 3.0 | 100 | 25 | [60] |
| **PV** | 703 | 395 | 340 | 326 | 1.0 | 100 | 25 | |
| **Battery** | 277 | 147 | 124 | 102 | 2.5 | 95 | 15 | [61] |
| **$LH_2$ storage** | 0.85 | 0.85 | 0.85 | 0.85 | 2.0 | 100 | 20 | [53] |
| **PEM electrolysis** | 900 | 700 | 575 | 450 | 1.5 | 2020:64<br>2030:69 | 19 | [62] |
| | | | | | | 2040:72<br>2050:74 | | |
| **Liquefaction** | 610 MEUR/GW^(0.66)*<br>(Plant SIZE)^-0.34 | | | | 1.5 | 82 | 20 | [53] |
| | **CAPEX [MEUR/km/GW]** | | | | | | | |
| **Electrical grid** | 0.9 | 0.9 | 0.9 | 0.9 | 1.5 | 99%<br>/1000km | 60 | [63] |
| **H2 pipeline** | 0.185 | 0.185 | 0.18<br>5 | 0.18<br>5 | 2.0 | 100 | 40 | [29] |
| | **CAPEX [MEUR/km]** | | | | | | | |
| **Local transmission grid (500MVA)** | 0.45 | 0.45 | 0.45 | 0.45 | 2.0 | 99%<br>/1000km | 60 | [63] |

### 2.3 Hydrogen cost-potential curve generation

Based on the derived hydrogen export costs as described above, the export cost-potential curves for liquid hydrogen are calculated. This is done for the years 2020, 2030, 2040, and 2050 based on the techno-economic assumptions from Table 2. For each year and country, nine different hydrogen export amounts are calculated to discretize the hydrogen cost-potential-curve. The export amounts are evenly spaced to 95% of the maximum exportable hydrogen. Above this threshold the optimization is limited in flexibility options like curtailment and storage capacities due to energy losses, which would lead to artificial system designs. By this, the most expensive 5% of the technical potential are neglected due to the effects above. The maximum is derived by applying the conversion efficiencies for electrolysis and liquefaction to the maximum electrical potential from Table 1. For each country, the combination of exported hydrogen amount and total system cost across all export variations form the cost-potential curve. In this study, from the 1008 possible configurations (28 countries, 9 export demands, 4 years), 957 combinations are calculated to derive a detailed cost-potential curve for each considered country. 51 energy systems achieved only suboptimality by the used optimization problem solver gurobi and, hence, were dropped. As these costs account for all energy system costs for each discrete assumed hydrogen export, the cost-potential curves are based on the absolute costs and not the marginal cost, which are often seen in economic evaluations of price models [64].

## 3 Results

This section presents the cost-potential curves for the liquid green hydrogen export of all considered countries. The countries are categorized into three groups based on different characteristics of their hydrogen export energy systems. Subsequently, these energy systems are analyzed in terms of their specific designs obtained within the optimization and their cost development through 2050 is shown.

### 3.1 Global green liquid hydrogen cost-potential curves

The obtained hydrogen cost-potential curves are shown Figure 5. In total, the exportable amount of liquid hydrogen for the considered 28 countries sums up to over 1540 $PWh_{LHV}$/a, which is about nine times the world primary energy consumption for 2019 (173 $PWh_{LHV}$) [65].

Investigating the resulting energy system designs in greater detail reveals three distinct country groups, as displayed in Figure 5.

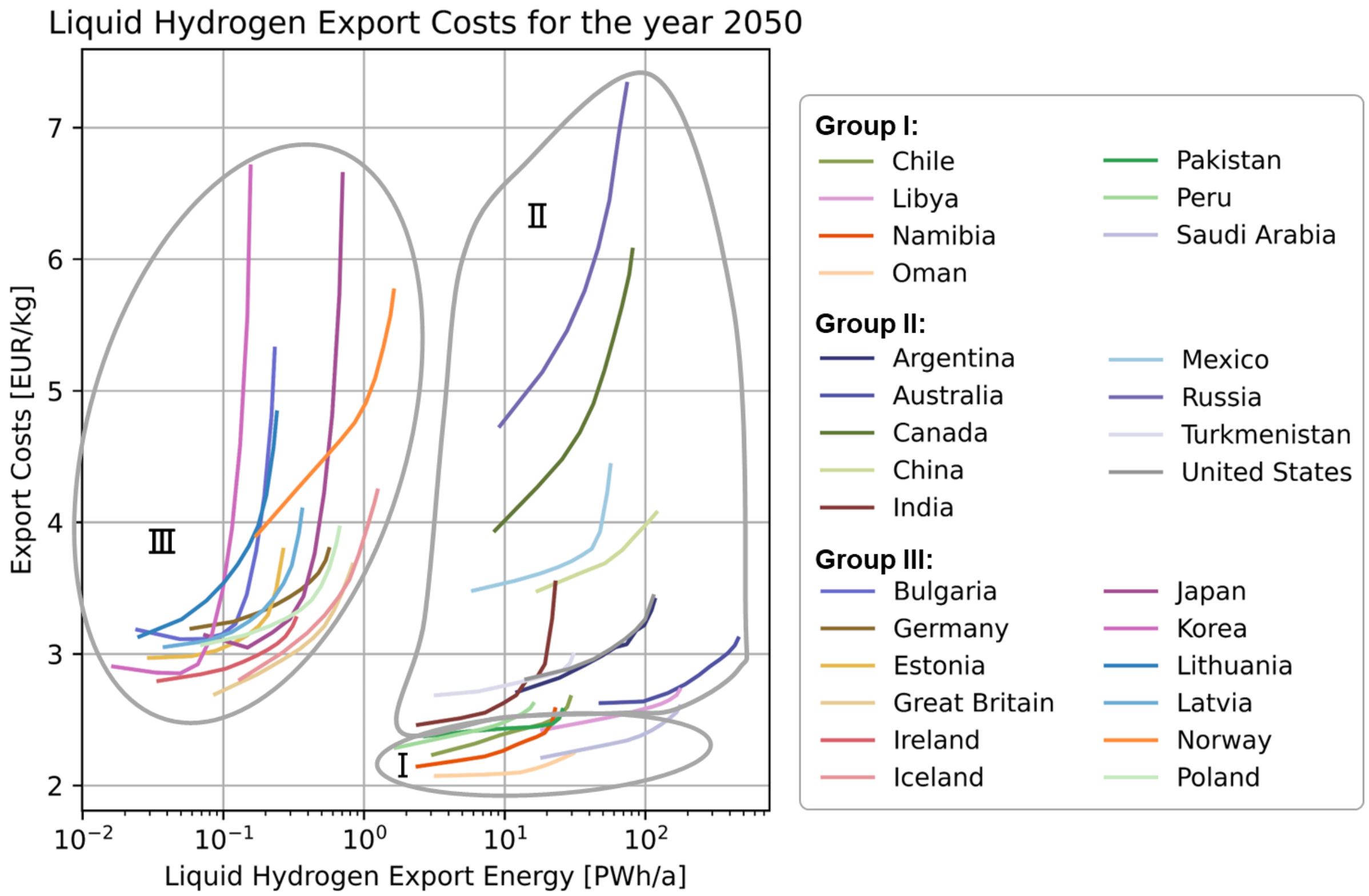

**Figure 5. Green liquid hydrogen export cost-potential curves for each considered country in 2050.**

Group I comprises countries with comparably cheap large-scale solar energy potentials. The hydrogen energy systems of those countries are dominated by solar energy, which leads to nearly stable hydrogen costs for the entire hydrogen potential (see the typical country of Oman in Figure 6). The small cost increase over the exported hydrogen stems mainly from transport, especially hydrogen grid costs, whose impact is constrained by decreasing liquefaction cost from scaling effects. The cost share of open-field PV and electrolysis as the main cost factors account for 53–65% for these countries. With high full load hours of open-field PV of up to and over 2000 h/a and minimal electricity generation costs of 1.63 EURct/kWh$_{el}$ in 2050 in Oman, a hydrogen cost of 2.07 EUR/kg$_{H2}$ can be achieved. In total, 79 PWh$_{LHV}$/a of liquid hydrogen for export can be produced at a hydrogen cost below 2.30 EUR/kg in 2050 within group I.

Group II includes the medium- to high-cost, large potential countries between 2.50 and 7,50 EUR/kg$_{H2}$. For some countries of this group, the effects of stable potential costs as described in group I also applies for the first part of their cost-potential curve (India, Turkmenistan and Australia with PV, and Argentina with wind). These parts sum up to 129 PWh$_{LHV}$/a hydrogen production and represent roughly 12% of the total potential of group II. Additionally, all of them exhibit a steep increase in their costs at a certain point (see Figure 6 for the typical country of Canada). This is primarily due to an increase of storage and grid costs with increasing exploitation of hydrogen potentials from 0.84 to 2.52 EUR/kg$_{H2}$ and an increase in electricity cost due to uneven distributed potentials with higher needs for comparably longer grid distances from 4.70 to 5.29 EURct/kWh$_{el}$.

Group III describes countries with smaller generation potentials below 1 $PWh_{LHV}$/a. As the 11 renewable energy clusters per GID-1 region are distributed over smaller potentials, more details can be observed. One example of this is the impact of scaling liquefaction costs, leading to a drop with increasing exploitation of the hydrogen potential for countries with steady potentials at low export rates such as South Korea and Bulgaria. In general, these countries see a substantial increase in costs with increasing hydrogen exports (see Figure 6 for South Korea) resulting from a reduction in wind full load hours from, e.g., over 2600 to below 1300 hours per year in South Korea. Iceland shows the same effect but resulting, in contrast, from constant high wind full load hours and low full load hours for PV placements. Only Germany and Ireland exhibit steady costs for green hydrogen because of their more constant RES full load hours. The distribution and storage costs exhibit a share of about 17% and are therefore the lowest of all the groups.

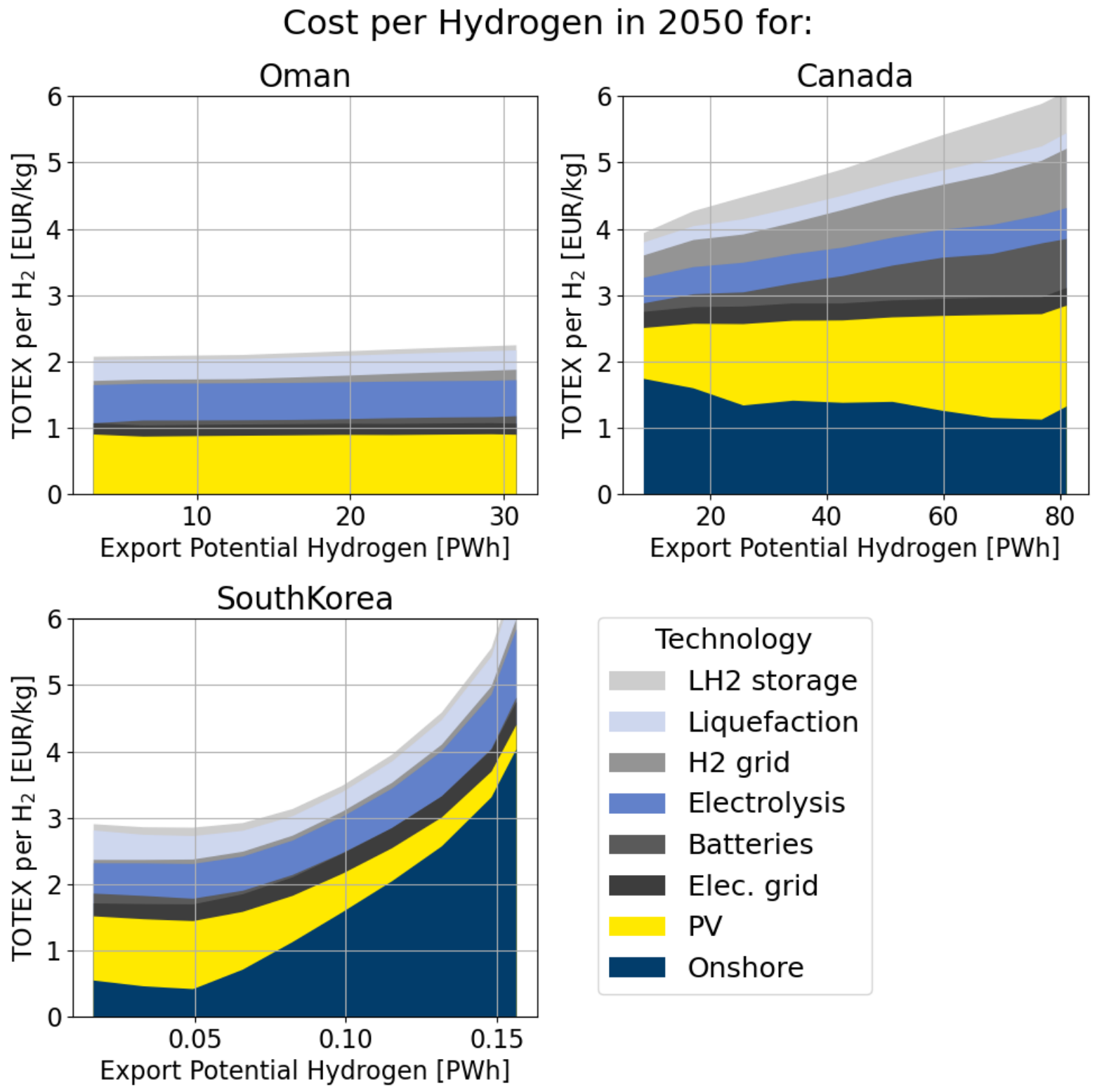

**Figure 6. Cost share per technology for the liquid hydrogen costs for typical countries of each group (Oman: group I; Canada: group II; South Korea: group III).**

### 3.2 Impact on optimal energy system design

Analyzing the resulting energy system designs in more detail reveals further patterns across the three country groups, as can be seen in Table 3. From the three available flexibility options (battery usage, grid balance, and curtailment) within the model, the option preferred by the model is curtailment of solar power combined with decentralized hydrogen production in the GID-1 regions for group I. The only countries utilizing batteries to a larger extent are from group II, resulting in a cost share of ca. 12% compared to less than 4% in the other country groups. In contrast, the larger countries of groups I and II require larger grid connections to exploit their full potentials stretched over their entire areas, resulting in a cost share for grids of roughly 18% and 24%, respectively, for groups I and II and only for 10% for the small countries of group III.

**Table 3. Technology cost per hydrogen generation and curtailment for each group.**

| Group | Total [EUR/kg] | Absolute costs [EUR/kg] | | | | | | | Curtailment |
|---|---|---|---|---|---|---|---|---|---|
| | | RES | PEM | Batteries | $LH_2$ tank | elec. grid | Pipelines | Liquefaction[1] | |
| I | 2.41 | 0.92 | 0.54 | 0.10 | 0.11 | 0.24 | 0.20 | 0.29 | 8.8% |
| II | 3.70 | 1.57 | 0.45 | 0.45 | 0.19 | 0.30 | 0.50 | 0.24 | 10.1% |
| III | 3.65 | 2.22 | 0.52 | 0.04 | 0.21 | 0.24 | 0.13 | 0.29 | 5.9% |

[1]: electricity costs for liquefaction are modelled endogenously and therefore are not separately included here but are included within the RES and grid costs (~0.2 EUR/kg at 3 EURct/$kWh_{el}$).

Although the interpretation of high grid costs for large countries is fairly trivial, the high utilization of batteries in group II derives from a combination of high transport costs due to long distances and comparably low full load hours of renewable energy technologies. This implies that the usage of batteries to decrease green hydrogen costs is only beneficial in comparable uneconomical hydrogen energy systems with hydrogen costs above 2.50 EUR/kg. Therefore, they will probably not be part of an optimal global solution for hydrogen supply, as other regions most likely offer a sufficient amount of potential green hydrogen. Moreover, the hydrogen energy systems from groups II and III typically utilize a combination of wind and solar power (see Figure 6) to make use of synergies in the different feed-in time series, ultimately leading to higher full load hours of the electrolysis (3140 h/a for group I and over 4000 h/a for groups II and III). Yet, the impact on the PEM cost reduction from the full load hour increase is minor, at only 0.02 to 0.09 EUR/$kg_{H2}$.

Figure 7 shows the impact of the share of PV utilization on the electricity generation costs. Firstly, it can be seen that all low-cost solutions (group I) only utilize solar energy primarily as a result of the higher costs of wind turbines. The average electricity generation cost in the considered regions results in roughly 0.062 EUR/$kWh_{el}$ for wind turbines and 0.026 EUR/$kWh_{el}$ for open-field PV. This difference mainly stems from the assumed cost of wind turbines and open-field PV and the different weather conditions. Secondly, all countries, even those with the best onshore wind placements in the world, utilize a higher capacity of PV than wind power in their hydrogen energy systems. This explains why some countries with high wind full load hours, such as Norway, still have high hydrogen generation costs due to comparably low full load hours of PV.

In addition, it must be noted that the lowest supporting points of the derived cost-potential curves in this study are at 10% of the maximum hydrogen export of each country, which already represent large-scale hydrogen production. Hence, smaller competitive wind energy potentials might be overlooked, which could contribute to local small-scale hydrogen production. However, the focus of this study is on the global exchange of hydrogen, for which large scale production units will be required to make use of scaling effects.

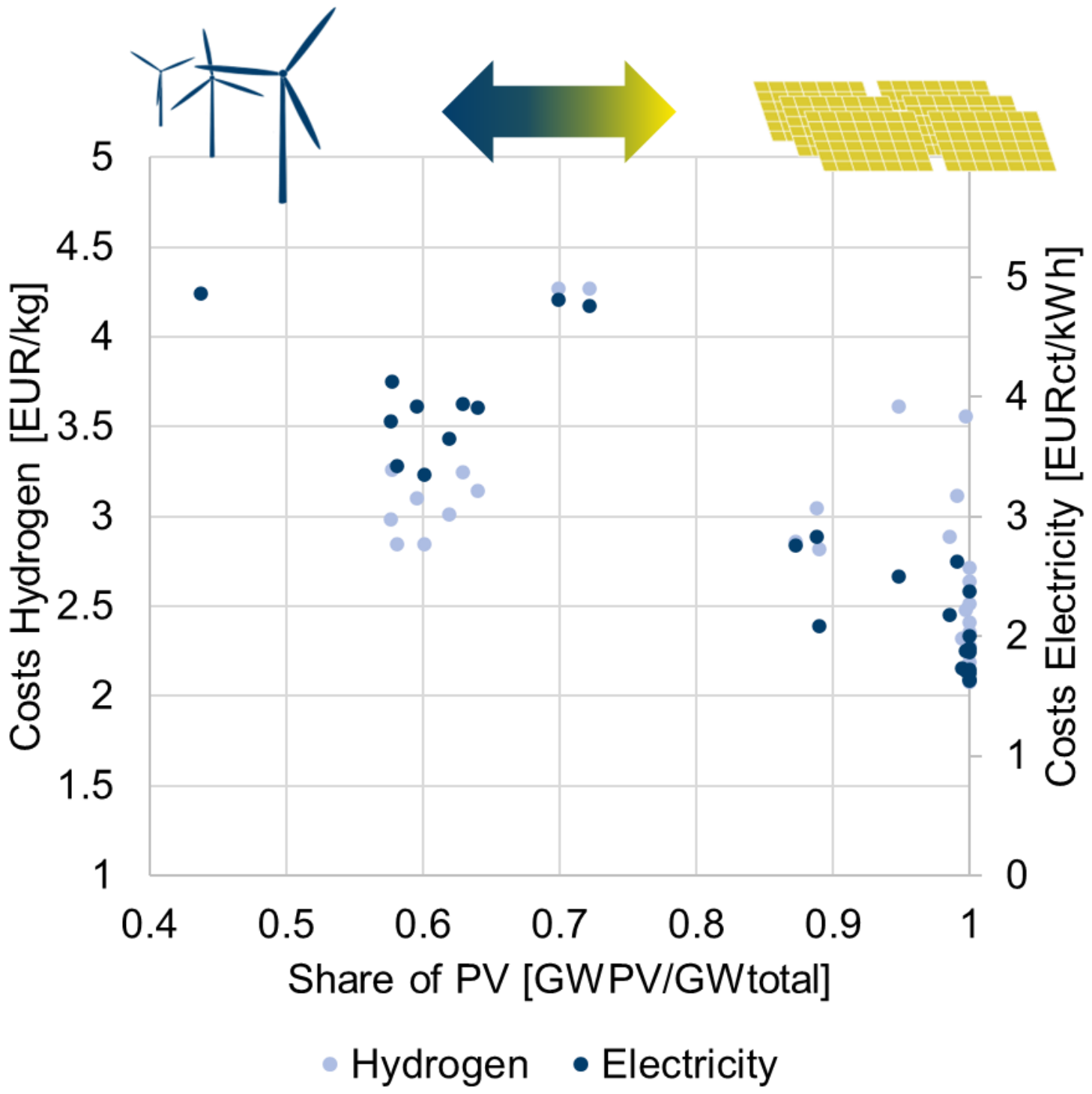

**Figure 7. Impact of PV and wind combinations on hydrogen production costs for each country at 20% export in 2050.**

## 3.3 Cost development

In this study, the optimal energy systems were derived for 2020 to 2050. The results are shown in Figure 8. It can be seen that in most cases, costs homogeneously drop. The only difference observed is between countries with rich solar and wind resources, where the average electricity costs between 2020 and 2050 drop by 52% for PV and 20% for onshore wind. This stems from assumptions regarding cost developments for onshore wind and solar PV, with solar countries experiencing a larger drop compared to wind countries between 2020 and 2030. Apart from that, the results do not indicate region-specific preferences for cost digressions for the observed countries. The biggest drop in costs for green liquid hydrogen of about 1.40 EUR/kg$_{H2}$ is expected to happen between 2020 and 2030. The first time that costs for green hydrogen production drop below 2.50 EUR/kg occurs in Oman and Namibia in 2040. Other countries only follow between 2040 and 2050 (see also Figure 5). It must be noted that these costs are still export costs in the harbor. Hence, the final hydrogen cost for local supply within the considered regions will not include the liquefaction cost, whereas imported hydrogen will bear additional shipping costs.

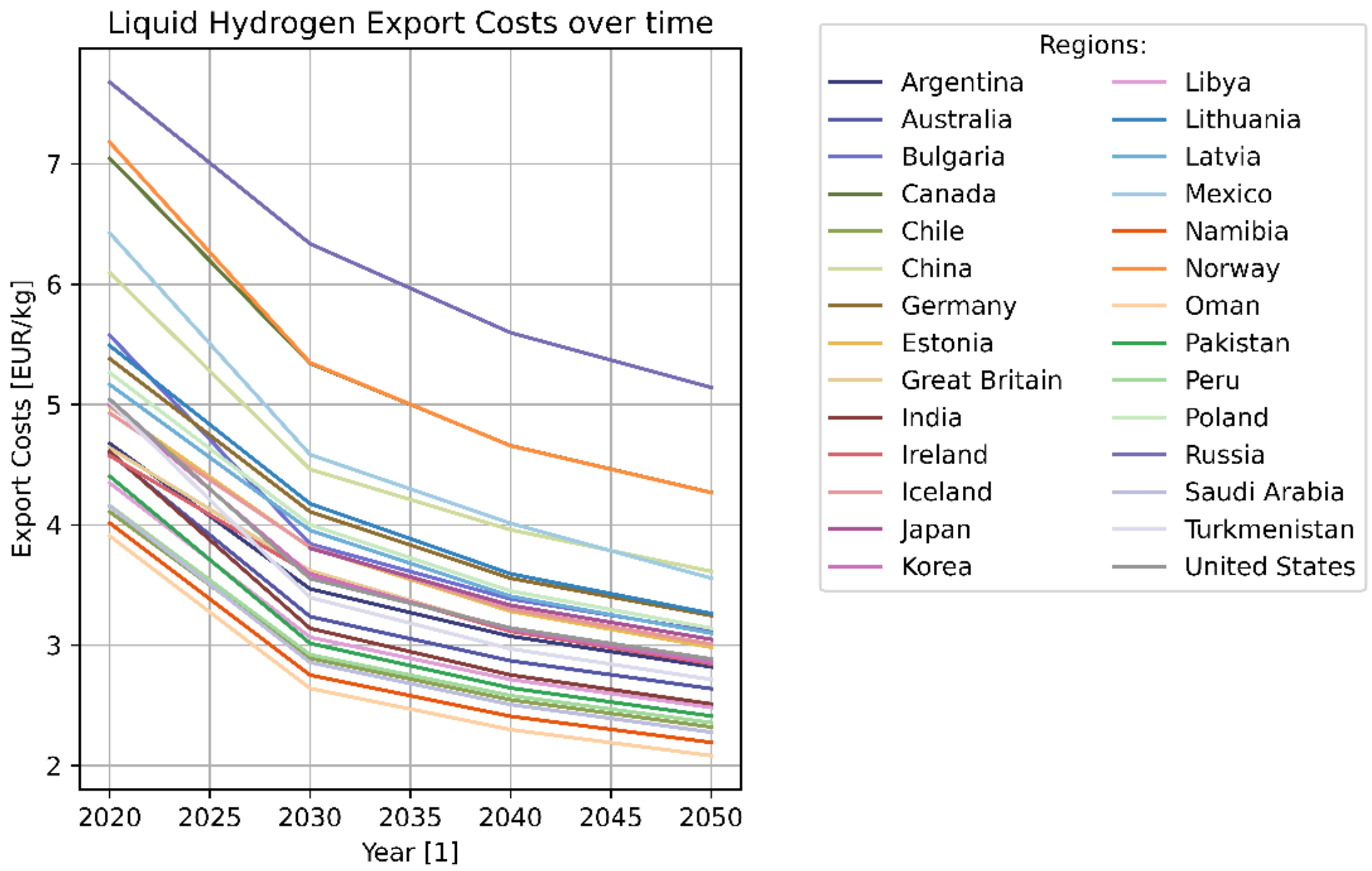

**Figure 8. Liquid hydrogen cost development from 2020 to 2050 at 20% of the maximum export.**

### 3.4 Cost Sensitivity Analysis

As the cost parameters for future technologies are forecasted, the assumptions from Table 2 underlie uncertainties. To tackle the impact of uncertainty in these costs a sensitivity analysis concerning the investment costs for PV, onshore wind, electrolysis, and liquefaction costs is conducted. The results for three exemplary countries, Oman for group I, Argentina for group II and Germany for group III are shown in Figure 9, where the investment cost for PV, onshore wind and PEM are varied by +-30% for an exemplary expansion rate of 10% of the country-wise maximum export in 2050. This expansion rate was chosen as it allows for most freedom in system design while already representing a global hydrogen market beyond a ramp-up. For the liquefaction, the maximum plant size is compared to a minimum of 700 t/d as a derived future plant capacity used by IRENA [66] and a theoretical unlimited scaling of the liquefaction costs.

There is a significant impact on liquid hydrogen export costs in Oman, particularly regarding liquefaction costs, which have an average impact of 19%. Oman's large export amounts result in the largest drop of hydrogen costs across all sensitivities, when infinite liquefaction scaling is applied, as the liquefaction plant is scaled from 20 kt/d to 808 kt/d. The corresponding cost are reduced by -0.22 EUR/kg$_{H2}$ to 1.85 EUR/kg$_{H2}$. However, Oman utilizes only solar resources which have lower full load hours than the wind resources of countries in groups II and III. As a result, limiting liquefaction to 700 t/d leads to the highest increase in liquefaction costs, making Oman and other countries in group I more sensitive to liquefaction costs than the other groups. Also, because Oman solely relies on solar resources, it shows the highest dependency on PV costs among all groups as seen in Table 4**.** Argentina shows the lowest dependency of all countries on specific changes in costs (below 11% for all sensitivities). This is due to the utilization of solar and wind resources, allowing the energy system to change the primary source of energy with changes in renewable costs without a significant change in electricity costs. In addition, the higher full load hours of the electrolysis and liquefaction due to the utilization of wind turbines make the energy system more robust to cost changes in electrolysis and liquefaction investments.

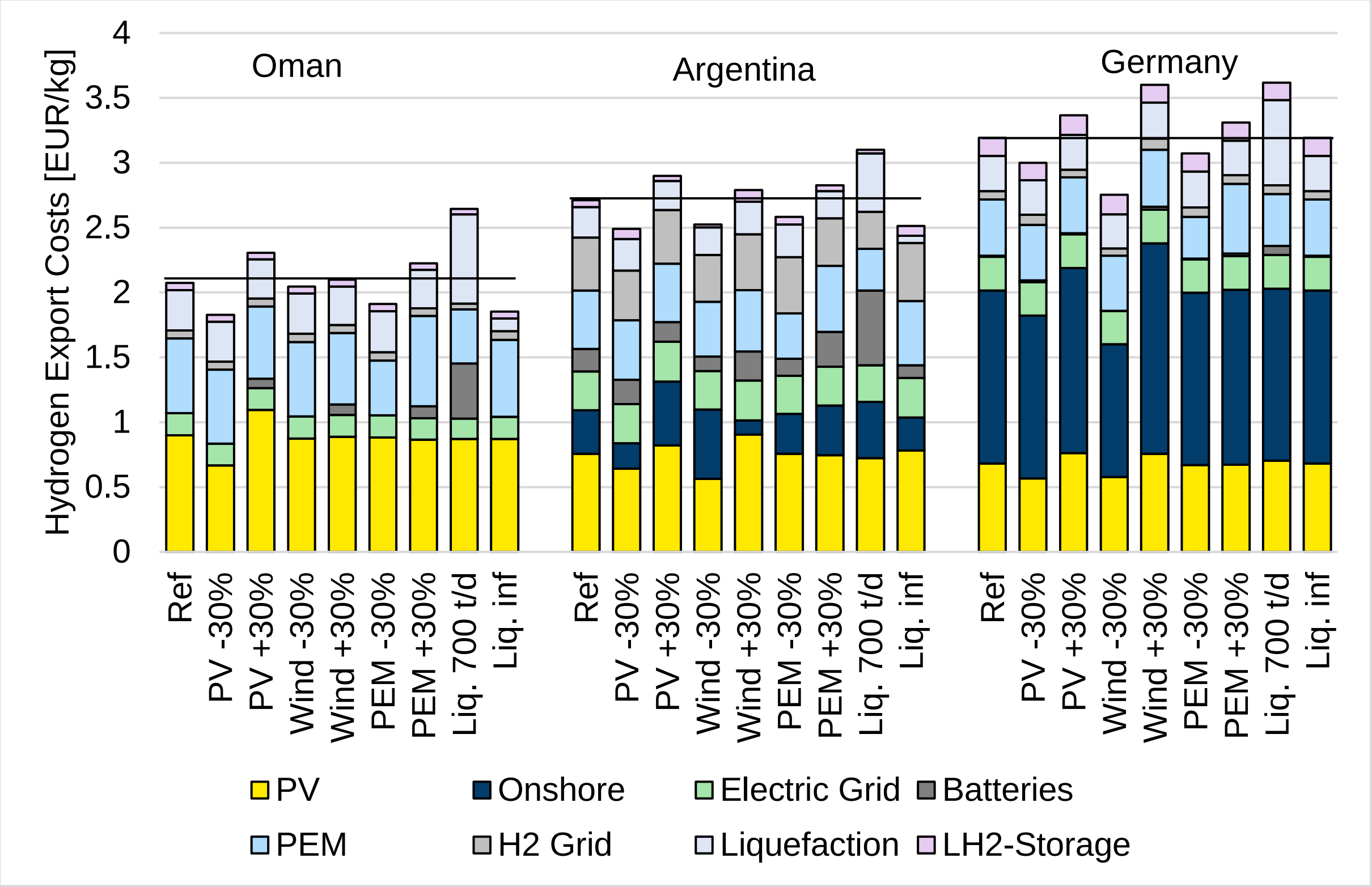

**Figure 9: Investment cost sensitivity analysis for Oman (group I), Argentina (group II) and Germany (group III) exemplarily for 10% of the country-wise maximum export in 2050.**

In Germany, the largest change in hydrogen costs occurs with varying onshore wind costs (13% change), as the energy system mainly utilizes onshore wind resources for electricity generation. Consequently, it shows the lowest change in hydrogen costs when PV costs change (6%). Because of the high full load hours of wind turbines in Germany, the change of electrolysis and liquefaction costs is low compared to the other countries, with the same reasoning as in Argentina. Furthermore, due to the small hydrogen export amount of Germany, as well as all countries in group III, infinite scaling of the liquefaction plant does not affect the hydrogen costs because the liquefaction plant is already below the maximum scaling of 20 kt/d in the reference case. Countries from group II and group III show lower dependency of hydrogen export cost on costs assumptions compared to countries from group I apart from the dependency due to high shares of wind in Germany. Another notable finding is that all energy systems tend to exhibit a high increase in batteries in the case of small and expensive liquefaction plants. This is due to the higher costs of liquefaction for smaller plant sizes, which favors higher full load hours of the liquefaction, which can be achieved by utilizing batteries. In the reference case without batteries, the full load hours of the liquefaction are at 2996 h/a, which rise to 4250 h/a in the case for higher liquefaction costs. In this case, the system purpose of batteries is balancing the diurnal fluctuations of electricity supply.

**Table 4: Average impact of sensitivities on liquid hydrogen export costs by technology for exemplary countries**

| | PV | Onshore wind | PEM | Liquefaction |
|---|---|---|---|---|
| **Oman (group I)** | 12% | 1% | 8% | 19% |
| **Argentina (group II)** | 8% | 5% | 5% | 11% |
| **Germany (group III)** | 6% | 13% | 4% | 7% |

### 3.5 Geostrategic security of hydrogen supply and water risk levels

Although the results presented before focus on cost and potentials, other aspects can impact future global hydrogen exchange. In this context, the diversification of supply and water risk levels in particular are important. As the considered export countries exhibit all types of political regimes, the question can arise of whether this might impact the security of supply. By sorting all cost-potential curves in accordance with the underlying political regime of each export country, cumulative curves revealing the impact of sourcing liquid hydrogen from different regime types becomes apparent. For this sorting, the State of Democracy map of Lauth et al. [67] is utilized in its most recent version from 2019. The map classifies political regimes into five types (namely working and deficient democracies, hybrid regimes, moderate and hard autocracies). The resulting cumulative cost-potential curves (see Figure 10) show that the export costs of autocratic countries are cheapest up to 500 $PWh_{LHV}$/a, whereas the export cost premiums of democratic countries account for roughly 7% at a hydrogen expansion equal to the global primary energy supply in 2019 [65]. However, differences in transport cost are likely to further increase this difference; especially for Germany and Europe more widely, this can increase the cost premium to ~20–24%. It must be noted that the assumed political situation might change until 2050, as this study only evaluates the political situation based on 2019.

In addition to the diversification aspects, the water supply for hydrogen must also be considered in terms of sustainability and security of supply. This is especially true, as water electrolysis consumes water instead of using it, as do many other types of water demands. Water stress scenarios from the World Resources Institute [68] reveal that even under the optimistic water stress scenario for 2040, most green hydrogen potential, and especially cheap solar-based hydrogen potential, is concentrated in regions that are already water-stressed, without imposing further burdens on water through hydrogen production (Figure 11). Water stress is assumed to start at the latest with a high water risk level [69].

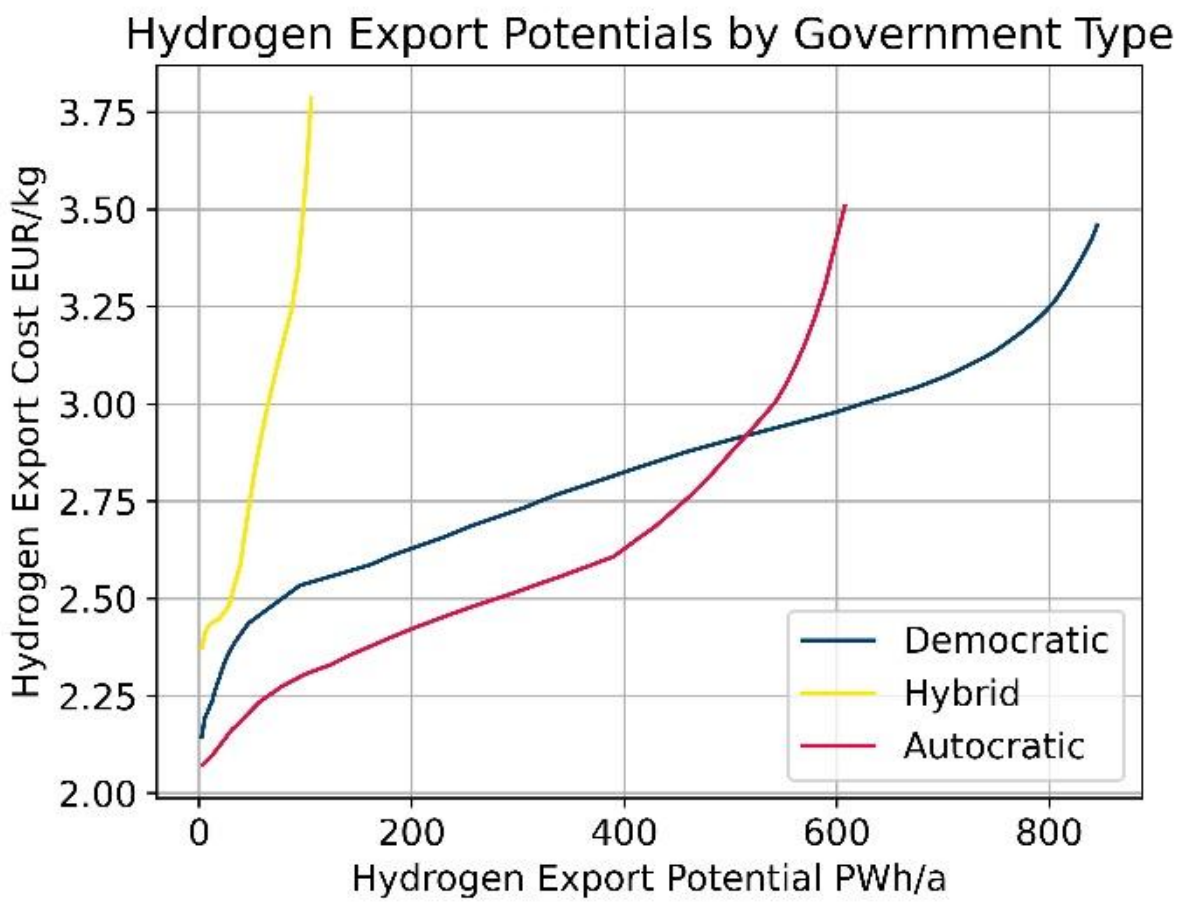

**Figure 10. Global hydrogen supply in 2050 by government type, based on [52].**

As each kg of hydrogen roughly requires 9 liters of water [50], the water demand for hydrogen production only in regions with high and extreme water risk alone adds up to more than 1.8 times Europe’s water withdrawal in 2019 [70]. However, utilizing seawater desalination in countries with sufficient coastal access can offer an alternative if installed in accordance with best practices so as to avoid environmental issues in coastal regions. Nevertheless, the impact of this additional process step on costs is only around 0.01 $EUR/kg_{H2}$, as a study by Heinrichs et al. [52] showed that utilizing sea water desalination for water for hydrogen energy systems adds little to no cost surplus to green hydrogen costs.

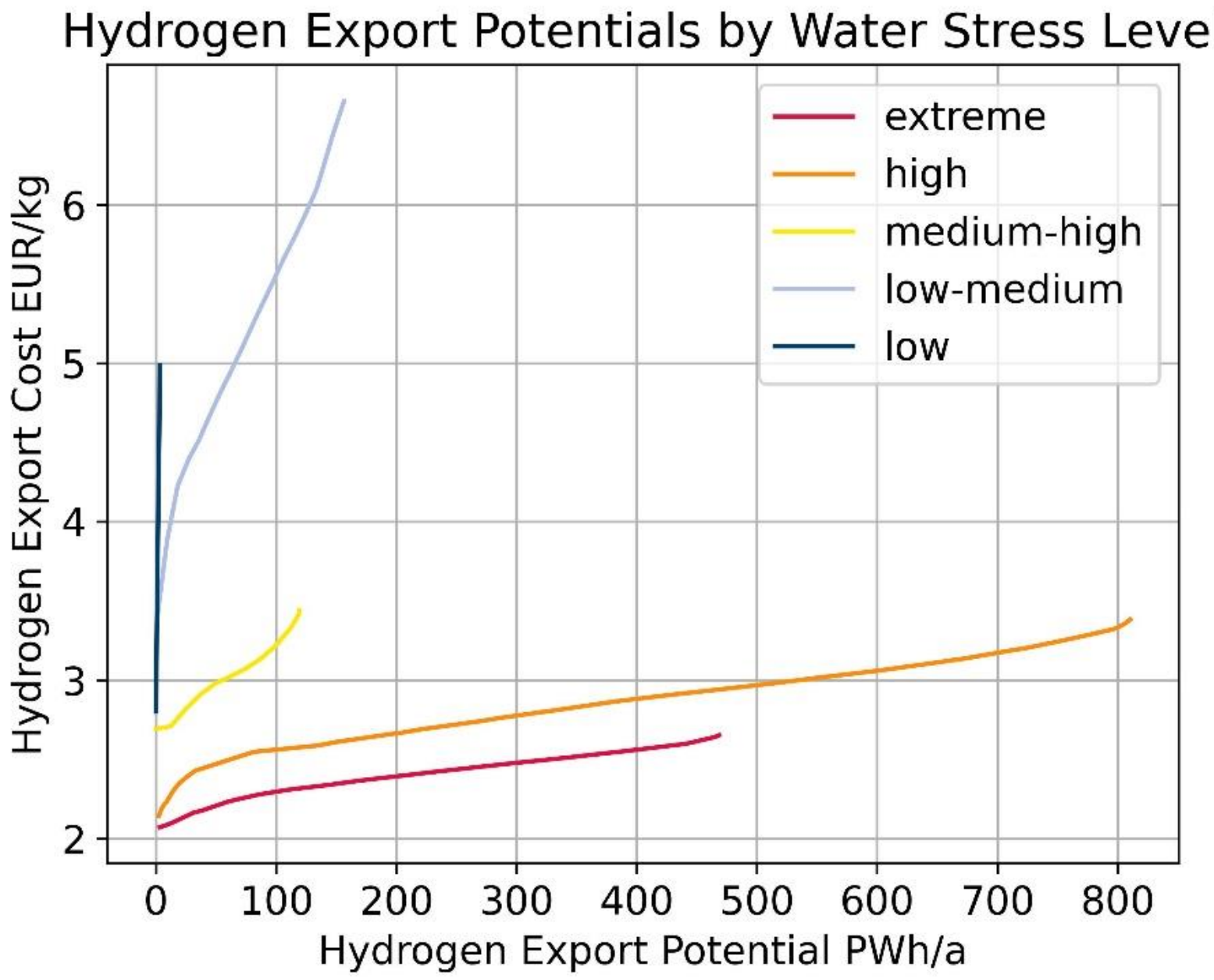

**Figure 11. Global hydrogen potentials in 2050 by water stress level, according to [54].**

## 4 Discussion and conclusions

The conducted analysis shows that the global green liquid hydrogen potentials of more than 1,540 $PWh_{LHV}/a$ in 28 countries exceed the world primary energy consumption by a factor of 9 [65]. Over 79 $PWh_{LHV}/a$ of this green liquid hydrogen will be available at costs below 2,30 EUR/kg in 2050. Therefrom, the highest cost contributors will be the renewable energy sources and electrolysis, amounting to about 65% of the total hydrogen costs. Hence, future hydrogen costs are highly sensitive to the decreasing costs of such technologies.

Compared to the results from IRENA [6], the costs reported in this study are generally higher. For example, in Australia, IRENA [6] estimates a cost of 0.8 EUR/kg in 2050, whereas this report projects over 2.6 EUR/kg. This is mainly explicable due differences in cost assumptions for CAPEX, WACC, and electrolysis efficiency (electrolysis: CAPEX = ~0.4 EUR/kg, efficiency = ~0.15 EUR/kg), as well as additionally considered energy system components in this study such as liquefaction, grids, and storage (~1.2 EUR/kg). Furthermore, Brändle et al. [21] calculates a lower cost for liquid hydrogen generation with, e.g., 1.8 EUR/kg in Australia for the same reason of not accounting for transportation and storage costs. Looking at hydrogen costs for Australia minus such costs for transportation and storage, the liquid hydrogen costs of this study would result in 1.74 EUR/kg, which is even below the estimates of Brändle et al. [21]. The study from Janssen et. al. [20] shows cost for gaseous hydrogen starting from 1.66

EUR/kg in Ireland in 2050, whereas this study projects 2.79 EUR/kg. As Janssen does not include the costs for a complex transport infrastructure, the hydrogen costs from this study for only RES and electrolysis account to 2.09 EUR/kg. The residual difference in costs is arising from higher assumptions of the lifetime of 30 years in Janssen et. al. [20] for onshore wind, PV and PEM resulting in lower total hydrogen costs. This shows the high impact of considering the full process chain in high temporal and spatial resolutions on the resulting hydrogen cost in addition to differences in cost assumptions.

For 2050, the IEA [71] states that the costs for hydrogen from natural gas with carbon capture and storage (CCS) technology will be between 1.15 and 2.02 EUR/kg based on a price assumption of 2 EUR/MWh for natural gas and for coal gasification with carbon capture and storage at about 2.12 to 2.4 EUR/kg. Compared to the costs for green hydrogen, starting at 2.07 EUR/kg in this study, it can be concluded that green hydrogen could become cost-competitive with hydrogen from coal with CCS in 2040, whereas hydrogen from natural gas with CCS heavily depends on the price developments of natural gas. Yet, another important point for both technologies is the availability of cheap and secure $CO_2$ storage, as well as acceptance of such storage capacities.

The lowest cost for hydrogen potentials will be available from PV-rich countries. Additionally, countries with good wind resources also depend on having good PV resources to achieve competitive hydrogen costs. This strong tendency towards PV-utilization facilitating low hydrogen costs can be seen in the literature [6], [21] as well. In accordance with this study, IRENA [6] found PV to be the dominant electricity source for green hydrogen over onshore wind sources. However, the results in this study go even further and suggest that even countries with high wind potentials will need to utilize at least 57% of the build capacity from PV to achieve low-cost green hydrogen production. Contrary to this, Janssen et. al [20] found, that the cheapest hydrogen in Europe can be produced from wind turbines and not PV. The main difference lies in the average full load hours for PV and onshore wind. This study calculates 1000 h/a for PV and 3575 h/a for onshore wind in Ireland including endogenously optimized curtailment, while Janssen et al. uses 788 for PV and 3942 h/a for onshore wind. This comparison indicates that there are tipping points in energy systems for favoring wind or solar generation for green hydrogen based on the cost prognosis of PV-modules, wind turbines and electrolysis as well as variations of RES input full load hours. As this study showed, capacity results for Oman and Germany are robust against costs variations, whereas Argentina shows a larger shift in capacity utilization with cost variations. Therefore, the implied tipping points in energy systems are country specific and not occurring in every country. Future work could be done in further examining these tipping points for green hydrogen production.

Moreover, for the PV-rich countries, the model actively chooses an optimal curtailment with direct and decentral electrolysis over storing and distributing electricity. The studies from IRENA [6] and Brändle et al. [21] show the same utilization of curtailment of renewable energy sources to increase the full load hours of electrolysis and, hence, decrease hydrogen costs. In contrast, this study finds a utilization of storage systems for wind-rich regions that has not been used in other studies [6], [21] due to their chosen approaches.

In general, the presented results are heavily dependent on the techno-economic assumptions from Table 2. The expansion of hydrogen- and renewable-based energy systems will foster decreasing costs due to learning effects. This learning effect can be endogenously modeled, as in Brändle et al. [21], but still depends on exogenous assumptions regarding expansion scenarios containing their own type of uncertainty. Although the focus of this study is on the impact of PV and onshore wind potentials, in some regions renewables like offshore wind and hydropower have an impact on hydrogen generation costs. Given the focus of this study on the large-scale export of hydrogen, all statements are only valid for such large-scale deployment of green liquid hydrogen generation. Based on the discretization of the models used, there might still be some small-scale applications for hydrogen generation that exhibit different characteristics, such as smaller wind farms with high full load hours with direct electrolysis utilization. Further, the local demand for hydrogen and electricity is not considered in this study, as only the exportable amount of hydrogen is calculated. While this will have an impact on some smaller countries from group 1 with an average electricity demand of 35% of the renewable potentials, the impact of local demands is significantly lower in larger countries of group 1 (<0.3%) or countries from group 2 (<3%) [72], as seen in Table 7.

Future green hydrogen will be needed to decarbonize the shipping and aviation sectors in particular, as well as the ammonia, methanol, and iron industries, as shown in Lippkau et al. [73]. Those demand centers are usually geographically separated from the sun-rich regions that can produce hydrogen at a low cost, as presented in this study. Based on the supply curves, global trading with sun-poor regions would be conceivable and cost-optimal [73]. $LH_2$ shipping would be a suitable technology for both long and short distances if gaseous pipeline transport is not an option [53]. In addition, $LH_2$ shipping could be possible starting at 0.52 EUR/(PJ 1000km) in 2030 and could decrease to 0.26 EU/(PJ 1000km) in 2100 [52]. The results of a TIMES-based energy system analysis with TIAM from Lippkau et al. [73] show that the main oil-exporting nations (e.g., Middle Eastern, Asian, African, and South American countries) could shift their economies towards a hydrogen basis. Yet, as stated in the analysis above, aspects of geostrategic security of supply should be taken into consideration.

## Acknowledgements

A major part of this work has been carried out within the framework of the ETSAP-DE project (03EI1032B) funded by the German Federal Ministry for Economic Affairs and Climate Action (BMWK). In addition, parts of this work have been carried out within the framework of the $H_2$ Atlas-Africa project (03EW0001) funded by the German Federal Ministry of Education and Research (BMBF).

## Author contributions

Conceptualization: DF, HH; methodology and model development: DF, HH; methodology – land eligibility: CW, DF, HH; validation: DF; formal analysis: DF, HH; investigation: DF; writing – original draft: DF, HH; writing – original draft renewable energy potentials: TA, PB, TH; writing – original draft discussion TIAM: FL, MB; writing – review and editing: all authors; visualization: DF; supervision: HH, JL, and DS. All authors have read and agreed to the published version of the manuscript.

References

[1] D. Stolten *et al., Neue Ziele auf alten Wegen? Strategien für eine treibhausgasneutrale Energieversorgung bis zum Jahr 2045*, 2022.

[2] D. G. Caglayan *et al.,* “Technical potential of salt caverns for hydrogen storage in Europe,” *International Journal of Hydrogen Energy*, vol. 45, no. 11, pp. 6793–6805, 2020, doi: 10.1016/j.ijhydene.2019.12.161.

[3] N. D. Pawar *et al.,* “Potential of green ammonia production in India,” *International Journal of Hydrogen Energy*, vol. 46, no. 54, pp. 27247–27267, 2021, doi: 10.1016/j.ijhydene.2021.05.203.

[4] P. Nikolaidis and A. Poullikkas, “A comparative overview of hydrogen production processes,” *Renewable and Sustainable Energy Reviews*, vol. 67, pp. 597–611, 2017, doi: 10.1016/j.rser.2016.09.044.
[5] P.-M. Heuser, D. S. Ryberg, T. Grube, M. Robinius, and D. Stolten, “Techno-economic analysis of a potential energy trading link between Patagonia and Japan based on CO2 free hydrogen,” *International Journal of Hydrogen Energy*, vol. 44, no. 25, pp. 12733–12747, 2019, doi: 10.1016/j.ijhydene.2018.12.156.
[6] IRENA, “Global Hydrogen Trade to Meet the 1.5°C Climate Goal: Green Hydrogen Cost and Potential,” 2022. [Online]. Available: https://www.irena.org/publications/2022/May/Global-hydrogen-trade-Cost
[7] N.-J. Alejandro and N. de Blasio, “MIGHTY: Model of International Green Hydrogen Trade,” *Belfer Center for Science and International Affairs, Harvard Kennedy School*, 2022. [Online]. Available: https://nrs.harvard.edu/URN-3:HUL.INSTREPOS:37373227
[8] C. Johnston, M. H. Ali Khan, R. Amal, R. Daiyan, and I. MacGill, “Shipping the sunshine: An open-source model for costing renewable hydrogen transport from Australia,” *International Journal of Hydrogen Energy*, vol. 47, no. 47, pp. 20362–20377, 2022, doi: 10.1016/j.ijhydene.2022.04.156.
[9] IRENA, *Global hydrogen trade to meet the 1.5°C climate goal: Part I – Trade outlook for 2050 and way forward*. Abu Dhabi: International Renewable Energy Agency, 2022.
[10] G. Glenk and S. Reichelstein, “Economics of converting renewable power to hydrogen,” *Nat Energy*, vol. 4, no. 3, pp. 216–222, 2019, doi: 10.1038/s41560-019-0326-1.
[11] P. Runge, C. Sölch, J. Albert, P. Wasserscheid, G. Zöttl, and V. Grimm, “Economic comparison of different electric fuels for energy scenarios in 2035,” *Applied Energy*, 233-234, pp. 1078–1093, 2019, doi: 10.1016/j.apenergy.2018.10.023.
[12] M. Drechsler, J. Egerer, M. Lange, F. Masurowski, J. Meyerhoff, and M. Oehlmann, “Efficient and equitable spatial allocation of renewable power plants at the country scale,” *Nat Energy*, vol. 2, no. 9, 2017, doi: 10.1038/nenergy.2017.124.
[13] R. Bhandari, “Green hydrogen production potential in West Africa – Case of Niger,” *Renewable Energy*, vol. 196, pp. 800–811, 2022, doi: 10.1016/j.renene.2022.07.052.
[14] G. K. Karayel, N. Javani, and I. Dincer, “Green hydrogen production potential for Turkey with solar energy,” *International Journal of Hydrogen Energy*, vol. 47, no. 45, pp. 19354–19364, 2022, doi: 10.1016/j.ijhydene.2021.10.240.
[15] R. C. Pietzcker, D. Stetter, S. Manger, and G. Luderer, “Using the sun to decarbonize the power sector: The economic potential of photovoltaics and concentrating solar power,” *Applied Energy*, vol. 135, pp. 704–720, 2014, doi: 10.1016/j.apenergy.2014.08.011.
[16] J. Bosch, I. Staffell, and A. D. Hawkes, “Temporally-explicit and spatially-resolved global onshore wind energy potentials,” *Energy*, vol. 131, pp. 207–217, 2017, doi: 10.1016/j.energy.2017.05.052.
[17] M. Fasihi and C. Breyer, “Baseload electricity and hydrogen supply based on hybrid PV-wind power plants,” *Journal of Cleaner Production*, vol. 243, p. 118466, 2020, doi: 10.1016/j.jclepro.2019.118466.
[18] L. Sens, Y. Piguel, U. Neuling, S. Timmerberg, K. Wilbrand, and M. Kaltschmitt, “Cost minimized hydrogen from solar and wind – Production and supply in the European catchment area,” *Energy Conversion and Management*, vol. 265, p. 115742, 2022, doi: 10.1016/j.enconman.2022.115742.
[19] M. Fasihi, R. Weiss, J. Savolainen, and C. Breyer, “Global potential of green ammonia based on hybrid PV-wind power plants,” *Applied Energy*, vol. 294, p. 116170, 2021, doi: 10.1016/j.apenergy.2020.116170.
[20] J. L. Janssen, M. Weeda, R. J. Detz, and B. van der Zwaan, “Country-specific cost projections for renewable hydrogen production through off-grid electricity systems,” *Applied Energy*, vol. 309, p. 118398, 2022, doi: 10.1016/j.apenergy.2021.118398.
[21] G. Brändle, M. Schönfisch, and S. Schulte, “Estimating long-term global supply costs for low-carbon hydrogen,” *Applied Energy*, vol. 302, p. 117481, 2021, doi: 10.1016/j.apenergy.2021.117481.

[22] M. Moritz, M. Schönfisch, and S. Schulte, "Estimating global production and supply costs for green hydrogen and hydrogen-based green energy commodities," *International Journal of Hydrogen Energy*, vol. 48, no. 25, pp. 9139–9154, 2023, doi: 10.1016/j.ijhydene.2022.12.046.
[23] P. Buchenberg *et al.,* "Global Potentials and Costs of Synfuels via Fischer-Tropsch Process," MDPI Energies, 2023.
[24] B. van der Zwaan, S. Lamboo, and F. Dalla Longa, "Timmermans' dream: An electricity and hydrogen partnership between Europe and North Africa," *Energy Policy*, vol. 159, p. 112613, 2021, doi: 10.1016/j.enpol.2021.112613.
[25] S. Guthrie, S. Giles, F. Dunkerley, H. Tabaqchali, and A. Harshfield, *The impact of ammonia emissions from agriculture on biodiversity*, 2018.
[26] I. B. Ocko and S. P. Hamburg, "Climate consequences of hydrogen emissions," *Atmos. Chem. Phys.*, vol. 22, no. 14, pp. 9349–9368, 2022, doi: 10.5194/acp-22-9349-2022.
[27] M. Markiewicz *et al.,* "Environmental and health impact assessment of Liquid Organic Hydrogen Carrier (LOHC) systems – challenges and preliminary results," *Energy Environ. Sci.*, vol. 8, no. 3, pp. 1035–1045, 2015, doi: 10.1039/C4EE03528C.
[28] X. Zhu, M. Li, and B. Liu, "Acute ammonia poisoning in dolly varden char (Salvelinus malma) and effect of methionine sulfoximine," *Fish & shellfish immunology*, vol. 101, pp. 198–204, 2020, doi: 10.1016/j.fsi.2020.03.068.
[29] Dilara Gulcin Caglayan, Heidi U. Heinrichs, Martin Robinius, and Detlef Stolten, *Robust design of a future 100% renewable european energy supply system with hydrogen infrastructure*, 2020.
[30] Union of Concerned Scientists, "Environmental Impacts of Hydroelectric Power," 2013. [Online]. Available: https://www.ucsusa.org/resources/environmental-impacts-hydroelectric-power
[31] R. L. Rowe, N. R. Street, and G. Taylor, "Identifying potential environmental impacts of large-scale deployment of dedicated bioenergy crops in the UK," *Renewable and Sustainable Energy Reviews*, vol. 13, no. 1, pp. 271–290, 2009, doi: 10.1016/j.rser.2007.07.008.
[32] V. Paolini, F. Petracchini, M. Segreto, L. Tomassetti, N. Naja, and A. Cecinato, "Environmental impact of biogas: A short review of current knowledge," *Journal of environmental science and health. Part A, Toxic/hazardous substances & environmental engineering*, vol. 53, no. 10, pp. 899–906, 2018, doi: 10.1080/10934529.2018.1459076.
[33] *FINE.* [Online]. Available: https://github.com/FZJ-IEK3-VSA/FINE.git
[34] L. Welder, D. Ryberg, L. Kotzur, T. Grube, M. Robinius, and D. Stolten, "Spatio-temporal optimization of a future energy system for power-to-hydrogen applications in Germany," *Energy*, vol. 158, pp. 1130–1149, 2018, doi: 10.1016/j.energy.2018.05.059.
[35] Global Wind Atlas 3.0, a free, web-based application developed, owned and operated by the Technical University of Denmark (DTU). The Global Wind Atlas 3.0 is released in partnership with the World Bank Group, utilizing data provided by Vortex, using funding provided by the Energy Sector Management Assistance Program (ESMAP). For additional information: https://globalwindatlas.info.
[36] Global Solar Atlas 2.0, a free, web-based application is developed and operated by the company Solargis s.r.o. on behalf of the World Bank Group, utilizing Solargis data, with funding provided by the Energy Sector Management Assistance Program (ESMAP). For additional information: https://globalsolaratlas.info.
[37] IRENA, *Renewable power generation costs in 2018*. Abu Dhabi: International Renewable Energy Agency, 2018.
[38] M. Kumar, "Social, Economic, and Environmental Impacts of Renewable Energy Resources," in *Wind Solar Hybrid Renewable Energy System*, K. Eloghene Okedu, A. Tahour, and A. Ghani Aissaou, Eds., Erscheinungsort nicht ermittelbar: IntechOpen, 2020.
[39] R. Loulou, "ETSAP-TIAM: the TIMES integrated assessment model. part II: mathematical formulation," *CMS*, vol. 5, 1-2, pp. 41–66, 2008, doi: 10.1007/s10287-007-0045-0.

[40] K. Siala, H. Houmy, and S. A. H. Rodriguez, *python Generator of REnewable Time series and mAps.* [Online]. Available: https://pygreta.readthedocs.io/en/latest/theory.html
[41] D. S. Ryberg, Z. Tulemat, D. Stolten, and M. Robinius, “Uniformly constrained land eligibility for onshore European wind power,” *Renewable Energy*, vol. 146, pp. 921–931, 2020, doi: 10.1016/j.renene.2019.06.127.
[42] T. He, P. Pachfule, H. Wu, Q. Xu, and P. Chen, “Hydrogen carriers,” *Nat Rev Mater*, vol. 1, no. 12, 2016, doi: 10.1038/natrevmats.2016.59.
[43] M. Niermann, S. Timmerberg, S. Drünert, and M. Kaltschmitt, “Liquid Organic Hydrogen Carriers and alternatives for international transport of renewable hydrogen,” *Renewable and Sustainable Energy Reviews*, vol. 135, p. 110171, 2021, doi: 10.1016/j.rser.2020.110171.
[44] L. Sens, U. Neuling, K. Wilbrand, and M. Kaltschmitt, “Conditioned hydrogen for a green hydrogen supply for heavy duty-vehicles in 2030 and 2050 – A techno-economic well-to-tank assessment of various supply chains,” *International Journal of Hydrogen Energy*, 2022, doi: 10.1016/j.ijhydene.2022.07.113.
[45] J. Hampp, M. Düren, and T. Brown, “Import options for chemical energy carriers from renewable sources to Germany,” *PloS one*, vol. 18, no. 2, e0262340, 2023, doi: 10.1371/journal.pone.0281380.
[46] S. Patil, L. Kotzur, and D. Stolten, “Advanced Spatial and Technological Aggregation Scheme for Energy System Models,” *Energies*, vol. 15, no. 24, p. 9517, 2022, doi: 10.3390/en15249517.
[47] *Marineinsight.* [Online]. Available: www.marineinsight.com (accessed: Feb. 6 2023).
[48] worldshipping, *Top 50 Ports.* [Online]. Available: https://www.worldshipping.org/top-50-ports (accessed: 2022).
[49] Igor Korobov, *The Caspian Region gas pipeline development prospects Eastern Corridor of Turkmen gas export.* [Online]. Available: https://www.polsoz.fu-berlin.de/polwiss/forschung/grundlagen/ffn/veranstaltungen/termine/archiv/pdfs_salzburg/Korobov.pdf
[50] R. R. Beswick, A. M. Oliveira, and Y. Yan, “Does the Green Hydrogen Economy Have a Water Problem?,” *ACS Energy Lett.*, vol. 6, no. 9, pp. 3167–3169, 2021, doi: 10.1021/acsenergylett.1c01375.
[51] J. Yates *et al.,* “Techno-economic Analysis of Hydrogen Electrolysis from Off-Grid Stand-Alone Photovoltaics Incorporating Uncertainty Analysis,” *Cell Reports Physical Science*, vol. 1, no. 10, p. 100209, 2020, doi: 10.1016/j.xcrp.2020.100209.
[52] H. Heinrichs, C. Winkler, D. Franzmann, J. Linssen, and D. Stolten, “Die Rolle von Meerwasserentsalzungsanlagen in einer globalen grünen Wasserstoffwirtschaft,” in 2021. [Online]. Available: https://juser.fz-juelich.de/record/905482
[53] M. Reuß, T. Grube, M. Robinius, P. Preuster, P. Wasserscheid, and D. Stolten, “Seasonal storage and alternative carriers: A flexible hydrogen supply chain model,” *Applied Energy*, vol. 200, pp. 290–302, 2017, doi: 10.1016/j.apenergy.2017.05.050.
[54] M. Reuß, P. Dimos, A. Léon, T. Grube, M. Robinius, and D. Stolten, “Hydrogen Road Transport Analysis in the Energy System: A Case Study for Germany through 2050,” *Energies*, vol. 14, no. 11, p. 3166, 2021, doi: 10.3390/en14113166.
[55] U. F. Cardella, *Large-scale hydrogen liquefaction under the aspect of economic viability: Dissertation*, 2018.
[56] S. Z. Al Ghafri *et al.,* “Hydrogen liquefaction: a review of the fundamental physics, engineering practice and future opportunities,” *Energy Environ. Sci.*, vol. 15, no. 7, pp. 2690–2731, 2022, doi: 10.1039/D2EE00099G.
[57] International Group of Liquefied Natural Gas Importer, *GIIGNL Annual Report: The LNG industry*, 2022. Accessed: Apr. 8 2023. [Online]. Available: https://giignl.org/wp-content/uploads/2022/05/GIIGNL2022_Annual_Report_May24.pdf
[58] Gourobi. [Online]. Available: https://www.gurobi.com/
[59] *Scipy.* [Online]. Available: https://scipy.org/
[60] R. C. Pietzcker, S. Osorio, and R. Rodrigues, “Tightening EU ETS targets in line with the European Green Deal: Impacts on the decarbonization of the EU power sector,” *Applied Energy*, vol. 293, p. 116914, 2021, doi: 10.1016/j.apenergy.2021.116914.

[61] National Renewable Energy Laboratory, *2021 Electricity ATB Technologies and Data Overview*, 2021. Accessed: 2022. [Online]. Available: https://atb.nrel.gov/electricity/2021/index

[62] International Energy Agency, *IEA G20 Hydrogen report*, 2020. Accessed: 2022. [Online]. Available: https://iea.blob.core.windows.net/assets/29b027e5-fefc-47df-aed0-456b1bb38844/IEA-The-Future-of-Hydrogen-Assumptions-Annex_CORR.pdf

[63] C. Moles *et al., Energy Technology Reference Indicator (ETRI) projections for 2010-2050*. Luxembourg: Publications Office, 2014.

[64] Y. He, M. Hildmann, F. Herzog, and G. Andersson, "Modeling the Merit Order Curve of the European Energy Exchange Power Market in Germany," *IEEE Trans. Power Syst.*, vol. 28, no. 3, pp. 3155–3164, 2013, doi: 10.1109/TPWRS.2013.2242497.

[65] OurWordInData, *energy production consumption.* [Online]. Available: https://ourworldindata.org/energy-production-consumption (accessed: 2022).

[66] IRENA, *IEA G20 Hydrogen report: Assumptions*, 2019. Accessed: Apr. 8 2023. [Online]. Available: https://iea.blob.core.windows.net/assets/a02a0c80-77b2-462e-a9d5-1099e0e572ce/IEA-The-Future-of-Hydrogen-Assumptions-Annex.pdf

[67] H.-J. Lauth, O. Schlenkrich, and L. Lemm, "Democracy Matrix (DeMaX) Version 3 goes online," 2019. [Online]. Available: https://www.demokratiematrix.de/fileadmin/Mediapool/PDFs/Report/DeMaX_Report_2019_Growing_Hybridity.pdf

[68] M. Luck and Landis,M. ,Gassert, F., *Aqueduct Water Stress Projections: Decadal projections of water supply and demand using CMIP5 GCMs*. Washington, DC, 2015.

[69] R. Hofste *et al.,* "Aqueduct 3.0: Updated Decision-Relevant Global Water Risk Indicators," *WRIPUB*, 2019, doi: 10.46830/writn.18.00146.

[70] The world bank, "Annual freshwater withdrawls, total (billion cubic meters) - European Union," 2023. Accessed: 2023. [Online]. Available: https://data.worldbank.org/indicator/ER.H2O.FWTL.K3?locations=EU

[71] International Energy Agency, "Global average levelised cost of hydrogen production by energy source and technology, 2019 and 2050," International Energy Agency, 2022. Accessed: 2023. [Online]. Available: https://www.iea.org/data-and-statistics/charts/global-average-levelised-cost-of-hydrogen-production-by-energy-source-and-technology-2019-and-2050

[72] EIA, *Data Browser.* [Online]. Available: https://www.eia.gov/opendata/browser

[73] F. Lippkau *et al.,* "Global Hydrogen and Synfuel Exchanges in an Emission free Energy System," *MDPI Energies*, 2023.

[74] Energy Information Agency, "Electricity: Electricity net consumption 2021," Energy Information Agency, 2022. Accessed: Apr. 11 2023. [Online]. Available: https://www.eia.gov/international/data/world/electricity/electricity-consumption?pd=2&p=0000002&u=0&f=A&v=mapbubble&a=-&i=none&vo=value&&t=C&g=00000000000000000000000000000000000000000000001&l=249-ruvvvvvfvtvnvv1vrvvvvfvvvvvvfvvvou20evvvvvvvvvvnvvvs0008&s=315532800000&e=1609459200000